\documentclass[twoside,fleqn]{article} 
\usepackage{espcrc2,epsfig}
\newcommand{\AmS}{{\protect\the\textfont2
    A\kern-.1667em\lower.5ex\hbox{M}\kern-.125emS}}

\title{\rightline{\small  HUB--EP--96/33}
  \vspace*{-3mm}
  \rightline{\small July 26, 1996}
  Detailed  Phase Transition Study at $M_H \leq 70$ GeV
  in a 3-dimensional $SU(2)$--Higgs Model }

\author{M.~G\"urtler$^1$\thanks{Contribution presented by M.~G\"urtler},
  E.-M.~Ilgenfritz$^2$, J.~Kripfganz$^3$, H.~Perlt$^1$ and
  A.~Schiller$^1$\\
  {\it $^1$ Institut f\"ur Theoretische Physik, Universit\"at Leipzig, Germany}\\
  {\it $^2$ Institut f\"ur Physik, Humboldt-Universit\"at zu Berlin, Germany}\\
  {\it $^3$ Institut f\"ur Theoretische Physik, Universit\"at Heidelberg,
    Germany} }

\begin{document}

\begin{abstract}
  We study the electroweak phase transition in an effective $3$--dimensional
  theory for a Higgs mass of about 70 GeV by Monte Carlo simulations.  The
  transition temperature and jumps of order parameters are obtained and
  extrapolated to the continuum using multi--histogram techniques and finite
  size analysis.
\end{abstract}

\maketitle
\section{Introduction}
\vspace*{-2mm} One approach to lattice calculations of the electroweak
transition is based on an effective $3$--dimensional $SU(2)$--Higgs model. It
is attractive phenomenologically because it circumvents the problem of putting
chiral fermions on the lattice. Due to dimensional reduction, fermions as well
as non--static bosonic modes contribute to the effective action. In contrast to
QCD, dimensional reduction should work for the electroweak theory around and
above the transition temperature because $g^2$ is small. For the electroweak
phase transition this approach has been pioneered by Farakos et al. (see {\it
  e.g.} \cite{Kajantieaug95}).  In its simplest version the dimensionally
reduced effective theory is again an $SU(2)$--Higgs theory with just one
doublet.  \vspace*{-2mm}
\section{The lattice model}
\vspace*{-2mm} On the lattice, we study the $SU(2)$--Higgs system with one
complex Higgs doublet of variable modulus. The gauge field is represented by
unitary $2 \times 2$ link matrices $U_{x,\alpha}$, the Higgs fields are written
as $\Phi_x = \rho_x V_x$ ($V_x \in SU(2)$).  The lattice action is
\begin{eqnarray}
  S &=& \beta_G \sum_p \big(1 - {1 \over 2} \mathrm{Tr}\ U_p \big)
  \nonumber \\
  & & - \beta_H \sum_l {1\over 2} \mathrm{Tr}\ (\Phi_x^+ U_{x, \alpha}
  \Phi_{x + \alpha}) \nonumber \\ 
  & & + \sum_x \big( \rho_x^2 + \beta_R (\rho_x^2-1)^2
  \big)
  \label{eq:latt_action}
\end{eqnarray}
($\rho_x^2= {\frac{1 }{2}} \mathrm{Tr}\ (\Phi_x^+\Phi_x)$, $U_p$ denotes the
$SU(2)$ plaquette matrix), with
\begin{eqnarray}
  \beta_G & = & {\frac{4 }{a g_3^2 }},\ \label{eq:betar}\\
  \beta_R & = & {\frac{\lambda_3 }{g_3^2}} \ {\frac{\beta_H^2
      }{\beta_G}}= {\frac 18 \left( \frac{M_H^*}{80 \mathrm{GeV}}\right)^2} 
  \ {\frac{\beta_H^2
      }{\beta_G}}, \\
  \beta_H & = & \frac{2 (1-2\beta_R)}{6+a^2 m_3^2}\ 
  \label{eq:betah}
\end{eqnarray}
($a$ is the lattice spacing).  The lattice model defined by
(\ref{eq:latt_action}) is numerically studied at given couplings $\beta_G,
\beta_H$ and $\lambda_3/g_3^2$.

In the search for the phase transition, bulk variables like
\begin{eqnarray}
  \label{eq:order_param}
  {\rho^2} & = & { 1 \over {L_x L_y L_z}} \sum_x \rho_x^2 ,\\
  {E_{link}} & = &  { 1 \over {3 L_x L_y L_z}} \sum_{x,\alpha} {1\over 2}
  \mathrm{Tr}\  (\Phi_x^+ U_{x, \alpha} \Phi_{x + \alpha}) 
\end{eqnarray}
are  used. 

The update is a combination of $3d$ and $4d$ Gaussian heat bath for the gauge
and Higgs fields, respectively, and Higgs reflections.  Most of the Monte Carlo
data have been obtained on QUADRICS parallel computers.

\vspace*{-2mm}
\section{Phase separation, equal weight and mixed phase configurations}
\vspace{-2mm} As usual, the search for the phase transition point requires
extensive application of the multi--histogram technique \cite{fs,BunkEA}.

We have studied the phase transition driven by $m_3$. Then the lattice Higgs
self--coupling $\beta_R$ varies with $\beta_H$ (see (\ref{eq:betar})).
Therefore, the reweighting uses not only ${E_{link}}$, but ${\rho^2}$ and
${\rho^4}$ at the same time.

We have determined the finite volume pseudo--critical $\beta_{Hc}(L)$ by the
minima of the Binder cumulants and the maxima of the susceptibilities
\begin{eqnarray}
  \label{eq:binder_cum} 
  B_{\rho^2}(L,\beta_H) & = & 1 - \frac{\big\langle {(\rho^2)}^4
    \big\rangle} {{\ 3 \big\langle {(\rho^2)}^2 \big\rangle}^2},
  \label{eq:suscept} \\
  C_{\rho^2}(L,\beta_H) & = & \big\langle {(\rho^2)} ^2
  \big\rangle  - {\big\langle {\rho^2} \big\rangle}^2
\end{eqnarray}
(in the same way for other observables) as well as using the equal 
weight method.

The application of the equal weight criterion requires a procedure to separate
the (measured or reweighted) histogram into contributions attributed to the
pure phases. In addition, there are inhomogeneous (mixed) configurations
contributing to the histogram.  Our main assumption is that the pure phases can
be described by Gaussian distributions for any volume averaged quantity.  The
normalised histogram has been represented as a weighted sum of three histograms
\begin{eqnarray}
  p({\rho^2},\beta_H)& =& w_b p_b({\rho^2},\beta_H) + w_s p_s({\rho^2},\beta_H)\nonumber \\
  & &  + w_{mix} p_{mix}({\rho^2},\beta_H)
  \label{eq:histog}
\end{eqnarray}
with $w_b+w_s+w_{mix}=1$.  $w_{b,s}$ denotes the weight of the broken/symmetric
phase, $w_{mix}$ that of the mixed state.

The positions, widths and weights of the pure phase histograms at a given
$\beta_H$ have been obtained by fitting the outer flanks of the two--peak
histogram to Gaussian shape. This fixes the weight $w_{mix}$ and the ${\rho^2}$
distribution to be attributed to configurations with domains of both phases in
equilibrium.  The pseudo--critical $\beta_{Hc}$ is found according to the
requirement $w_b = w_s$.

Another kind of phase separation was used to estimate the jump of the plaquette
and correlation lengths of ''pure'' phases at the critical point.  The aim is
to remove successful tunneling escapes and unsuccessful tunneling attempts
towards the ''wrong'' phase from what should then be considered as the Monte
Carlo trajectory restricted to the ''right'' phase.

The procedure rescans the records of ${\rho^2}$ which has a well separated
two--peak signal for all considered volumes. Referring to this variable a lower
cut for the upper (broken) phase and an upper cut for the lower (symmetric)
phase can be chosen.  The cuts are determined in such a way that the remaining
histograms (for the ''pure'' phases) are almost symmetric around their maxima.
If the Monte Carlo history of ${\rho^2}$ enters the range of a certain phase
and stays there for a number of iterations (larger than the autocorrelation
time {\it without} tunneling but smaller than that {\it with} tunneling), the
sequence of configurations is considered to belong to that phase until the
trajectory leaves it.

\section{ Phase Transition Localisation}
\begin{figure}[thb]
  \begin{flushright}
    \epsfig{file=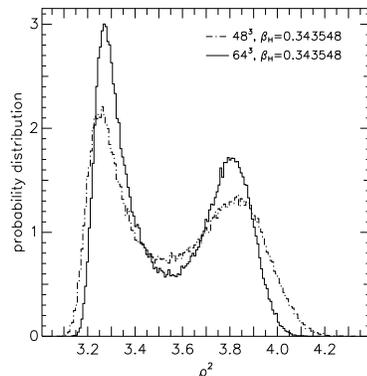,angle=0,height=50mm} \hspace*{10mm}
  \end{flushright}
  \vspace{-15mm}
  \caption{\sl Measured histograms of ${\rho^2}$ for $\beta_G=12$}
  \label{fig:rho2hist}
\end{figure}

To demonstrate the two--peak structure we show in Fig.~\ref{fig:rho2hist} the
measured histogram of ${\rho^2}$ on lattices $48^3$ and $64^3$, all at
$\beta_G=12$, for $\beta_H$ values nearest to the respective pseudo--critical
$\beta_{Hc}(L)$. The positions of the maxima change already only slightly with
the volume.

\begin{figure}[thb]
  \unitlength1mm
  \begin{flushright}
    \epsfig{file=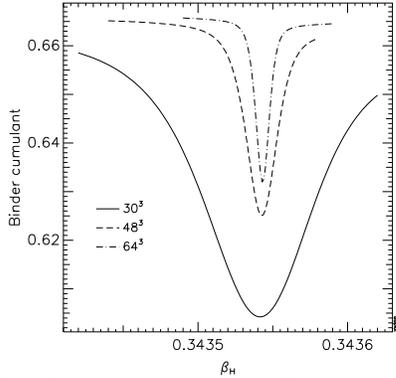,angle=90,height=50mm}
    \hspace*{10mm}
  \end{flushright}
  \vspace{-15mm}
  \caption{\sl Binder cumulant for {$E_{link}$}, $\beta_G=12$}
  \label{fig:fsana}
\end{figure}

In Fig.~\ref{fig:fsana} results of the multi--histogram interpolation of our
data for $\beta_G=12$ for the Binder cumulant of ${E_{link}}$ are presented.
Finiteness and shrinking of the Binder cumulant with increasing volume present
evidence for the first order nature of the transition at Higgs mass $M_H^*=70$
GeV.

\begin{figure}[thb]
  \begin{flushright}
    \epsfig{file=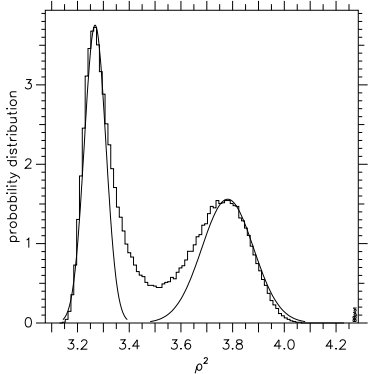,width=12cm,angle=0,height=50mm}
    \hspace*{10mm}
  \end{flushright}
  \vspace{-15mm}
  \caption{\sl Equal weight histogram}
  \label{fig:hist_rho2_equalweight}
\end{figure}

An equal--weight histogram ($\beta_{Hc}(L)=0.3435441$) for a lattice size
$64^3$ is presented in Fig.~\ref{fig:hist_rho2_equalweight} together with the
Gaussians describing the pure phases. The distribution attributed to mixed
configurations with domains of both phases in equilibrium is well identified
between the two peaks.

\begin{figure}[h]
  \unitlength1mm
  \begin{flushright}
    \epsfig{file=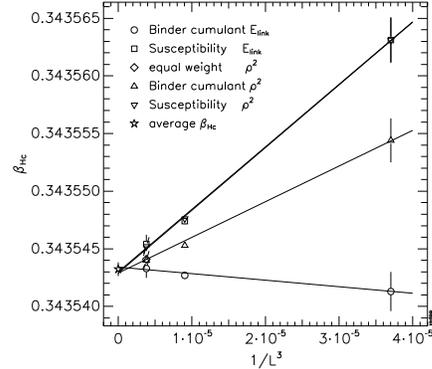,angle=90,height=50mm}
    \hspace*{10mm}
  \end{flushright}
  \vspace{-15mm}
  \caption{\sl Infinite volume extrapolation of $\beta_{Hc}$($\beta_G=12$)}
  \label{fig:beta_Hc}
\end{figure}

The various pseudo--critical $\beta_{Hc}(L)$ values for the different methods
applied to ${E_{link}}$ and ${\rho^2}$ are collected in Fig.~\ref{fig:beta_Hc},
plotted versus $1/L^3$ for $\beta_G=12$.  Corresponding to each method, a
$1/L^3$ fit has been used to yield a respective $\beta_{Hc}^{\infty}$. The
extrapolations nicely coincide as expected.

In Table~1 the extrapolations for each method are collected together with the
average $\beta_{Hc}^{\infty}$ for $\beta_G=12 $ and $16$. The result is
translated into a physical temperature and an "exact" Higgs mass $M_H$ using
the correspondence to quantities in the $4$--dimensional theory in $1$--loop
accuracy (no $\mu_4$-dependence, see \cite{Kajantieaug95}).

These numbers are given for the case of the $SU(2)$--Higgs theory without and
with fermions ({\sl i.e.} including the top quark with a mass of $175$ GeV).
Comparing the temperatures, there seems to be not much space left for $O(a)$
corrections. The "exact" Higgs mass is practically the same.
\begin{table}[thb]
  \begin{center}
    \begin{tabular}{|@{}c|rrr@{}|}
      \hline
      $\beta_G=12$
      & $B$ & $C$ & $w_b=w_s$ \\
      \hline
      $E_{link}$  & 0.3435434 & 0.3435430 &               \\
      $\rho^2$   & 0.3435429 & 0.3435429 & 0.3435441      \\
      \hline
      $\overline {\beta_{Hc}}$ & \multicolumn{3}{c|}{0.3435433(6)} \\
      \hline
      \hline
      & no fermions  &   \multicolumn{2}{c|}{$m_t$=$175$\mbox GeV}\\
      \hline
      $T_c/ \mbox{GeV}$        & 150.94(1) &\multicolumn{2}{c|}{107.05(1)}   \\
      $M_H/ \mbox{GeV}$        & 64.77     & \multicolumn{2}{c|}{69.42} \\
      \hline
    \end{tabular}
    \vskip3mm
    \begin{tabular*}{75mm}{@{\extracolsep{\fill}}|c|rr|}
      \hline
      $\beta_G=16$ & $B$ & $C$ \\
      \hline
      $E_{link}$& 0.3407950 & 0.3407937         \\
      $\rho^2$  & 0.3407943 & 0.3407939               \\
      \hline
      $\overline {\beta_{Hc}}$ &  \multicolumn{2}{c|}{0.3407942(6)}  \\
      \hline
      \hline
      & no fermions  &   {$m_t$=$175$\mbox GeV}\\
      \hline
      $T_c/ \mbox{GeV}$        & 151.27(1) &{107.17(1)}   \\
      $M_H/ \mbox{GeV}$        & 64.77     &{69.46} \\
      \hline
    \end{tabular*}
    Table~1. \sl Infinite volume limit for $\beta_{Hc}$ at $M_H^*=70\ \mbox
    GeV$
    \label{tab:betah_infty}
  \end{center}
  \vspace*{-5mm}
\end{table}

For comparison, at the smaller coupling ($M_H^*=35$ GeV) the transition
temperature (without top) is $T_c=76.2(1)$ GeV with the Higgs mass $M_H=29.5$
GeV.  This has been obtained for gauge couplings in the range from $\beta_G=12$
to $20$ on lattices of size $40^3$ and $20^3$ \cite{physlett}.
\section{Condensate discontinuities}
\vspace{-2mm} The jumps in $\langle {\rho^2} \rangle$ and $\langle {\rho^4}
\rangle$ are connected to the renormalization group invariant discontinuities
of the quadratic and quartic Higgs condensates. The two--state signal for
$\langle {\rho^2} \rangle$ and $\langle {\rho^4} \rangle$ is still clearly
visible for all lattice sizes considered at the higher Higgs mass of $M_H^*=70$
GeV, where the transition turns out much weaker than at $M_H^*=35$ GeV.

\begin{figure}[thb]
  \unitlength1mm
  \begin{flushright}
    \epsfig{file=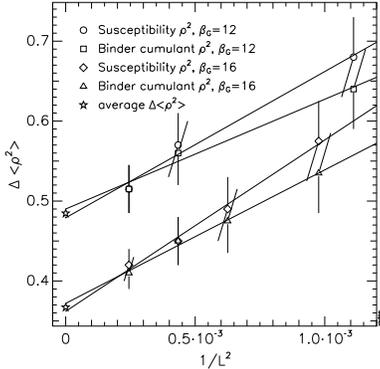,angle=90,height=50mm}
    \hspace*{10mm}
  \end{flushright}
  \vspace{-15mm}
  \caption{\sl Infinite volume extrapolation of $\Delta \langle
    {\rho^2} \rangle$ and $\Delta \langle {\rho^4} \rangle$}
  \label{fig:jump_rho}
\end{figure}

Corresponding to the different criteria applied for the definition of the
pseudo--critical $\beta_{Hc}(L)$ we obtain histograms of the various operators
just at the respective pseudo--criticality.  A collection of discontinuities of
$\langle {\rho^2} \rangle$ for various finite lattices (read off from the
maxima of the corresponding histograms) is shown in Fig.~\ref{fig:jump_rho}.
\vspace*{-2mm}
As in the analysis of Ref.~\cite{Kajantieoct95} we found that the size
dependence of the jumps for all available lattice sizes is best described by a
$1/L^2$ fit. The extrapolation to infinite volume is given in Table~2 in
lattice units.  \vspace*{-4mm}
\begin{table}[thb]
  \centering
  \begin{tabular*}{75mm}{@{\extracolsep{\fill}}|c|rrr|}
    \hline
    & $B_{\rho^2}$ &   $C_{\rho^2}$    &   average      \\
    \hline
    $\beta_G=12$  & 0.490(9)  &  0.479(8)   & 0.485(6)             \\
    $\beta_G=16$  & 0.372(8) & 0.362(7)  &    0.367(5)             \\
    \hline
    \hline
    & $B_{\rho^4}$   & $C_{\rho^4}$ &   average      \\
    \hline
    $\beta_G=12$  & 4.86(9)     &    4.81(8)        &    4.84(6)             \\
    $\beta_G=16$  & 3.62(8)     &    3.51(7)        &    3.56(5)             \\
    \hline
  \end{tabular*}
  Table~2. \sl Infinite volume limit for
  $\Delta \langle {\rho^2} 
  \rangle$ (upper) and $\Delta \langle {\rho^4} 
  \rangle$ (lower part),  \mbox{$M_H^*=70$ GeV}
  \label{tab:jumps_at_70}
\end{table}

Additionally, the expectation value of the average plaquette $\langle P
\rangle$ shows a discontinuity as well. This jump is numerically a tiny effect
at the larger Higgs mass.  Nevertheless, we are able to estimate it using the
phase separation technique discussed earlier.

In Table~3 the jump $\Delta \langle P \rangle$ is reported for $\beta_H$ values
nearest to the critical ones at $M_H^*=70$ and $35$ GeV at lattice sizes $64^3$
and $40^3$, respectively.
\begin{table}[!h]
  \vspace{-5mm}
  \centering
  \begin{tabular*}{75mm}{@{\extracolsep{\fill}}|rr|r|}
    \hline
    & & $\Delta \langle P \rangle$ \\
    \hline
    $M_H^*=70$ GeV & $\beta_G=12$ & 0.00037 \\
    &              $\beta_G=16$ & 0.00015\\
    \hline
    $M_H^*=35$ GeV & $\beta_G=12$ & 0.00370\\
    \hline
  \end{tabular*}
  Table~3. \sl Estimated plaquette jump $\Delta \langle P
  \rangle$
  \label{tab:jump_plaq}
  \vspace*{-2mm}
\end{table}

The relation of the measured quantities to continuum physics as well as further
consequences are discussed in the related contribution of A. Schiller
\cite{arwed}.  \vspace*{-3mm}
\section{Summary}
\vspace*{-2mm} We have studied the electroweak phase transition for Higgs
masses up to $70$ GeV. Its first order nature has been demonstrated. The
critical parameters were determined using the multi--histogram technique in
conjunction with the equal weight and other more standard criteria giving
consistent results.  The jumps in several quantities are reported and
extrapolated to the continuum.   
\end{document}